\documentclass[aps,pre,twocolumn]{revtex4-2}

\pdfoutput=1

\usepackage{mathtools}
\usepackage{amssymb}
\usepackage{xcolor}
\usepackage[colorlinks, citecolor={blue!70!black}, urlcolor={blue!70!black}, linkcolor={red!70!black},hyperindex,breaklinks]{hyperref}

\newcommand{\beq}{\begin{equation}}
\newcommand{\eeq}{\end{equation}}
\renewcommand{\hbar}{\mathchar'26\mkern-9mu h}
\newcommand{\w}{\omega}

\newcommand{\A}{\mathcal{A}}

\DeclareMathOperator{\Tr}{Tr}

\newcommand{\pii}{\mathit{\Pi}}

\begin{document}

\title{Microcanonical Truncations of Observables in Quantum Chaotic Systems}

\author{Fernando Iniguez}
\email{finiguez@ucsb.edu}
\author{Mark Srednicki}
\email{mark@physics.ucsb.edu}
\affiliation{Department of Physics, University of California, Santa Barbara, CA 93106}

\date{\today}

\begin{abstract}
We consider the properties of an observable (such as a single spin component that squares
to the identity) when expressed as a matrix in the basis of energy eigenstates, and then truncated
to a microcanonical slice of energies of varying width. For a quantum chaotic system, we model
the unitary or orthogonal matrix that relates the spin basis to the energy basis as a random matrix selected
from the appropriate Haar measure. We find that the spectrum of eigenvalues is given by a centered
Jacobi distribution that approaches the Wigner semicircle of a random hermitian matrix for
small slices. For slices that contain more than half the states, there is a set of eigenvalues
of exactly $\pm 1$. The transition to this qualitatively different behavior at half size is
similar to that seen in other quantities such as entanglement entropy. Our results serve as a
benchmark model for numerical calculations in realistic physical systems.
\end{abstract}

\maketitle
\section{Introduction}
The eigenstate thermalization hypothesis (ETH) \cite{Deutsch_91_Quantum,Srednicki_94_Chaos,Srednicki_96_Thermal,Srednicki_99_Approach,
Rigol_08_Thermalization,DAlessio_16_From,Deutsch_18_Review} 
is now widely accepted as a microscopic mechanism that is able to explain 
how an isolated quantum many-body system can come to thermal equilibrium
when starting from an initial pure state that appears to be far from equilibrium.
ETH is expected to hold for a ``chaotic'' quantum system that is sufficiently far (in a parameter space
of possible hamiltonians) from any point of integrability and which also does not exhibit many-body
localization due to strong disorder.
ETH then takes the form of an ansatz for the matrix elements (in the energy-eigenstate basis) of each
observable $A$ that would be measured in order to determine whether or not the system is in 
thermal equilibrium. This ansatz is
\begin{equation}
    A_{ij} = \A(E)\delta_{ij} + e^{-S(E)/2}f(E,\w)R_{ij},
    \label{eth}
\end{equation}
where $E=(E_i+E_j)/2$ is the average energy of the two eigenstates, $\w=E_i-E_j$ is their energy difference, $S(E)$ is the thermodynamic entropy (logarithm of the density of states) at energy $E$, 
$\A(E)$ and $f(E,\w)$ are smooth, real functions of their arguments, with $f(E,\w)=f(E,-\w)$, 
and $R_{ij}$ varies erratically, with overall zero mean and unit variance in local ranges of $E$ and $\w$.

A question of interest is whether more can be said about the statistical properties of the $R_{ij}$'s.
An argument based on the central limit theorem would indicate that they can be treated as independent
gaussian random variables, and numerical investigations in specific systems have generally been
consistent with this. However, as has been pointed out before \cite{Foini_19_ETH-OTOC,Richter_20_ETH_beyond}, 
this gaussianity cannot be an exact property, as it would yield various unphysical predictions, including an expression of any $n$-point time correlation function of $A$ in terms of the 2-point function. 
Furthermore, the operator $A$ has a spectrum of eigenvalues, and this spectrum must somehow 
be encoded in the energy-basis matrix elements $A_{ij}$. Because of this, 
as noted in \cite{Foini_19_ETH-OTOC}, 
it is more useful to think of the unitary matrix $U_{ai}$ that transforms basis states 
in which $A$ is diagonal to the energy-eigenstate basis as a statistically random matrix.

In \cite{Richter_20_ETH_beyond} (see also \cite{Pappalardi_23_WOW}), 
the observable $A$ was taken to be a component of a single spin in a lattice spin
system, and the eigenvalues of $A_{ij}$ computed when $i$ and $j$ were restricted to particular ranges of 
energies. If this submatrix had the statistical properties of a gaussian random matrix,
then a Wigner semi-circular distribution of eigenvalues would be expected. This was found for
small energy ranges, but significant deviations appeared at larger ranges. 

Our goal here is to provide a theoretical benchmark for these calculations, computing the expected
eigenvalue spectrum for a single-spin-component operator $A$ (which obeys $A^2=I$) when its matrix 
$A_{ij}$ in the energy-eigenstate basis is truncated, with the energies $E_i$ and $E_j$ each in the same
finite range. We refer to this as a microcanonical slicing. 
We specialize to the case where the ${\cal A}(E)$ function in Eq.~(\ref{eth}) is zero; this is equivalent to
\beq
\Tr e^{-\beta H}\!A = 0
\label{laSra}
\eeq
for all inverse temperatures $\beta$. This corresponds to a system in which the hamiltonian
$H$ is invariant under $A \to -A$. 
We then treat the diagonalizing matrix $U$ as either a unitary or orthogonal Haar-random matrix.
(The result is the same in both cases.)
This is the strongest possible assumption of random-matrix behavior; 
for an actual physical system, we expect correlations that result in $U$ having an 
approximately banded structure. We hope that our benchmark results can be used in the future
to help elucidate this structure in different physical systems of interest.

\section{Setup}

We consider an operator $A$ that obeys $A^2=I$ and $\Tr A =0$.
We take the dimension of the full Hilbert space to be $2D$; for a system of
$N$ two-component spins, we would have $2D=2^N$.
The eigenvalues of $A$ are hence $\pm 1$, with $D$ eigenvalues of each sign.
 
We then write
\beq
A = U^\dagger \tilde A U,
\label{UAU}
\eeq
where $\tilde A$ is a $2D\times 2D$ diagonal matrix,
\beq
\tilde A = 
\begin{pmatrix}
+I & 0 \\
\noalign{\smallskip}
0 & -I 
\end{pmatrix},
\eeq
and $U$ is a unitary (or orthogonal) matrix that transforms from the computational basis 
(in which a component of each spin is diagonal) to the
energy basis (in which the hamiltonian is diagonal). 

We are now interested in a microcanonical slicing of $A$, defined as
\beq
A_K = \pii_K A \pii_K
\label{AK}
\eeq
where $\pii_K$ is a projection operator 
onto an energy window that spans $K$ energy eigenstates.

We first specialize to the case $K\le D$.
We treat $U$ as either a unitary or orthogonal matrix that is selected at random from the 
corresponding Haar measure,
and consider the expected distribution $\rho(\lambda)$ of the eigenvalues $\lambda$ of $A_K$ 
in the limit of large $D$.

This problem has been solved in a different context \cite{Collins_05_Jacobi}, 
and the result (for either unitary or orthogonal $U$) is a special case of the centered Jacobi distribution,
\beq
\rho(\lambda) = \frac{\sqrt{4\alpha(1-\alpha)-\lambda^2}}{2\pi\alpha(1-\lambda^2)},
\label{rholam}
\eeq
where we have defined
\beq
\alpha = \frac{K}{2D},
\label{a}
\eeq
and where $\rho(\lambda)$ vanishes 
for values of $\lambda$ for which the argument of the square-root is negative. We have normalized $\rho(\lambda)$ to integrate to one. 

For a thin microcanonical slice, $K\ll D$ and hence $\alpha\ll 1$, the maximum value of
$\lambda^2$ is $4\alpha\ll 1$, and then we can replace $1-\lambda^2$ in the denominator
of Eq.~(\ref{rholam}) with $1$. We then have
\beq
\rho_{\alpha\ll 1}(\lambda) \simeq \frac{1}{2\pi\alpha}\sqrt{4\alpha-\lambda^2}.
\label{rholam2}
\eeq
We can compare this result with the Wigner semicircle for an $K\times K$ hermitian random matrix $H$
with matrix elements $H_{ij}$ drawn from a gaussian distribution with the expected value of $|H_{ij}|^2$,
$i\ne j$, given by $v^2$; this is \cite{Mehta_04_Random}
\beq
\rho_{\mathrm W}(\lambda) = \frac{1}{2\pi Kv^2}\sqrt{4Kv^2-\lambda^2}.
\label{rholamW}
\eeq
Our $2D\times 2D$ matrix $A$ obeys $A^2=I$, and hence 
\beq
\sum_{j=1}^{2D}|A_{ij}|^2=1.
\label{sumj}
\eeq
This implies that the expected value of each $|A_{ij}|^2$ is $1/2D$. 
If we set $v^2=1/2D$ in Eq.~(\ref{rholamW}),
we find that Eq.~(\ref{rholam2}) matches it.
This result agrees with the expectation from ETH that, for small energy differences, 
the energy-basis matrix elements of a local observable should have the statistics
of independent gaussian random variables.

For larger values of $\alpha$, Eq.~(\ref{rholam}) begins to differ from the Wigner semicircle.
The curvature at the origin $\rho''(0)$ is one diagnostic; it is negative for 
$\alpha<\alpha_{\mathrm c}=(2-\sqrt{2})/4=0.146$ but turns positive for $\alpha>\alpha_{\mathrm c}$. 
At $\alpha=1/2$, $K=D$, we find an arcsine distribution
with integrable singularities at $\lambda=\pm 1$,
\beq
\rho_{\alpha=1/2}(\lambda) = \frac{1}{\pi\sqrt{1-\lambda^2}}.
\label{rholam12}
\eeq

In Fig.~1(a,b,c), we show the eigenvalue distribution for a matrix with $2D=10{,}000$ 
and with $U$ a particular orthogonal matrix selected at random from the 
Haar measure, and with $K/2D = \alpha = 1/8, 1/4, 1/2$, along with the predicted
distribution of Eq.~(\ref{rholam}). We find very good agreement.

Next we consider the case $K\ge D$. This can be related to the case $K\le D$ by considering
the complementary microcanonical projection operator,
\beq
\pii_{2D-K} = I-\pii_{K}.
\label{piik}
\eeq
We show in the Appendix that for $K>D$ the eigenvalues of $A_K$ are minus those of $A_{2D-K}$,
plus $K-D$ extra pairs of eigenvalues of exactly $\pm 1$. This is true for any specific individual $U$. 
Hence, after averaging $U$ over a Haar measure, 
the result will be a continuous spectrum given by Eq.~(\ref{rholam}) (though with the normalizing
factor of $\alpha$ in the denominator replaced by $1-\alpha$), plus a discrete spectrum of 
$K-D$ eigenvalues $+1$ and $K-D$ eigenvalues $-1$.

In Fig.~1(d,e), we show the eigenvalue distribution for $K/2D = \alpha = 3/4, 7/8$.
We find, as predicted, a continuous distribution that matches that of the matrix
with $K/2D = 1-\alpha$, with all remaining eigenvalues exactly equal to $\pm 1$.

These results for $K>D$ are contrary to our initial expectations. 
We expected to find the Wigner semicircle
for $K\ll D$, and for this to gradually morph into a set of only $\pm 1$ at $K=2D$. Our expectations
are met by the results for $K\le D$, but the sudden appearance of some exact $\pm 1$ eigenvalues for
every $K>D$, along with an additional continuous distribution that mirrors the distribution for
$K<D$ and eventually becomes a Wigner semicircle for small $2D-K$, came as a surprise to us.
We will discuss this further in our conclusions.

\section{Conclusions}

Motivated by the investigations of \cite{Richter_20_ETH_beyond},
we have considered the properties of a single spin-component operator 
(with eigenvalues that are an equal number of plus ones and minus ones)
in a many-body quantum-chaotic system. We are interested in the statistical properties
of the matrix elements of such an operator in the energy-eigenstate basis.
We model this by treating the unitary (or orthogonal) matrix $U$ that relates
the spin-eigenstate basis to the energy-eigenstate basis as a random matrix
selected from the Haar measure. We then consider microcanonical truncations
of this matrix in the energy basis, and study their eigenvalues. For truncations
to a much smaller matrix, we find the distribution agrees with the Wigner
semicircle expected for a hermitian random matrix. For larger truncations,
we find that the truncated matrix begins to ``remember'' that the eigenvalues
of the full matrix are $\pm 1$.

Once the truncation is to a matrix larger than half the size of the original,
we find that there are now a set of eigenvalues of exactly $\pm 1$, along
with a continuous distribution that matches that of the complementary truncation.
This result was counter to what we initially expected, and shows a kind of
phase transition at half system size. This is reminiscent of a similar transition
in the behavior of entanglement and R\'enyi entropies \cite{Lu_19_Renyi},
which also exhibit sudden changes of behavior at half system size.
Similar transitions have also been discussed in \cite{Garrison_18_Single} 
for operators that include more than half the degrees of freedom of the system.

Our results are based on the most chaotic possible behavior of a physical system,
in which the energy eigenstates are completely random superpositions of basis states,
without any additional structure. Though this is unlikely to be true for any
realistic physical system, our results serve as a useful benchmark of comparison for
numerical calculations in these systems.

\begin{acknowledgments}
The work of F.I. was supported by an NSF Graduate Research Fellowship under Grant No.~2139319 and funds from the University of California. M.S. thanks Lauri Foini, Jorge Kurchan, and Silvia Pappalardi for helpful
discussions.
\end{acknowledgments}

\clearpage

\onecolumngrid
\begin{center}
\begin{figure*}
\begin{tabular}{cc}
  \includegraphics[width=85mm]{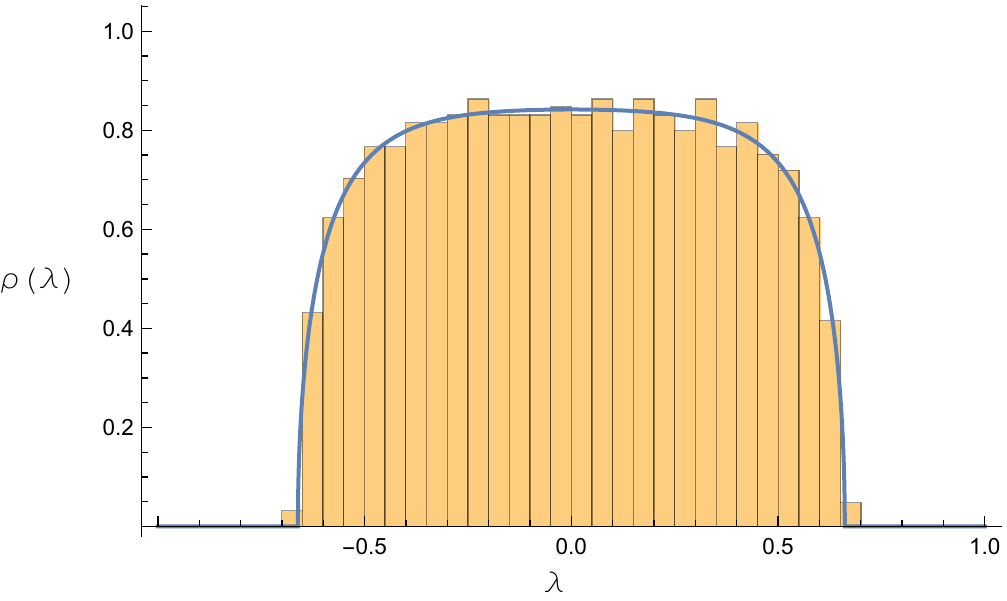} &   \includegraphics[width=85mm]{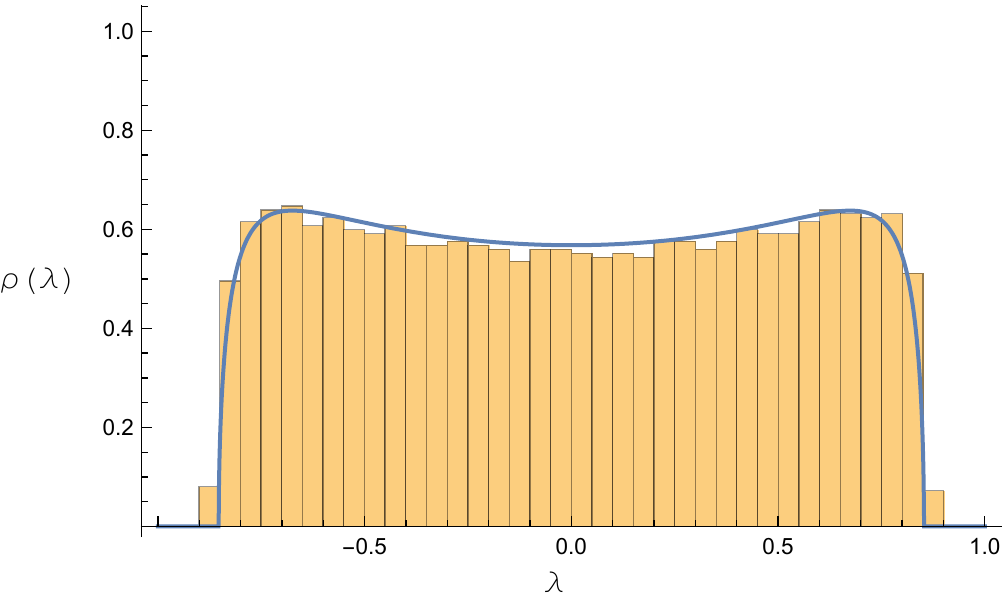} \\
(a)  & (b)  \\[6pt]
 \includegraphics[width=85mm]{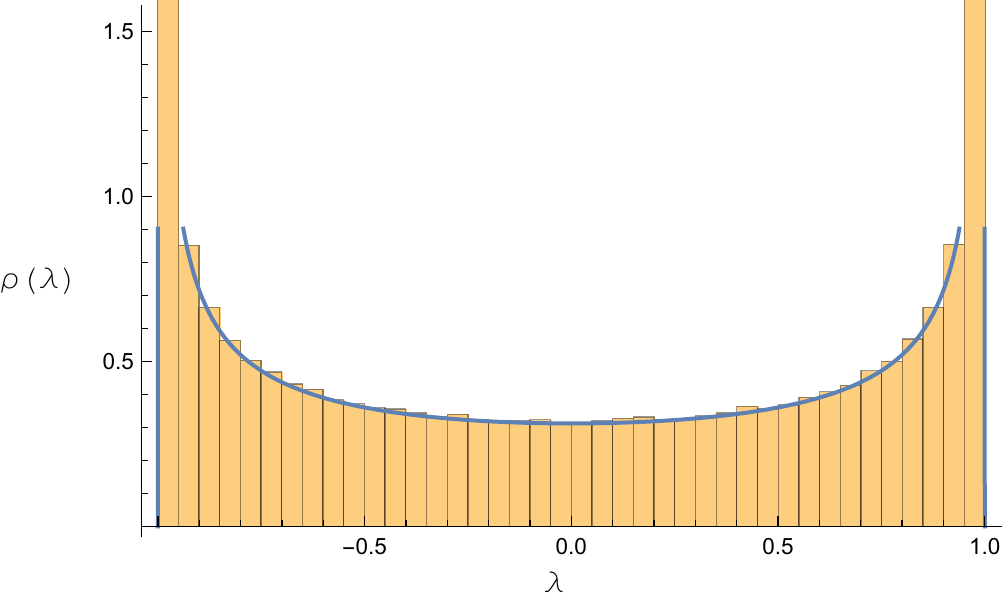} &   \includegraphics[width=85mm]{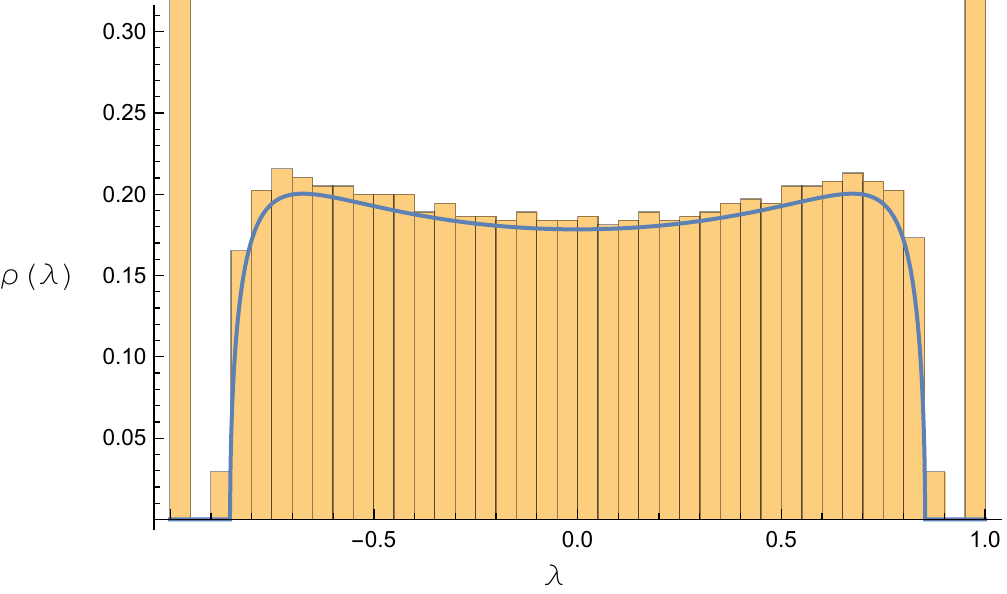} \\
(c) & (d)  \\[6pt]
\multicolumn{2}{c}{\includegraphics[width=85mm]{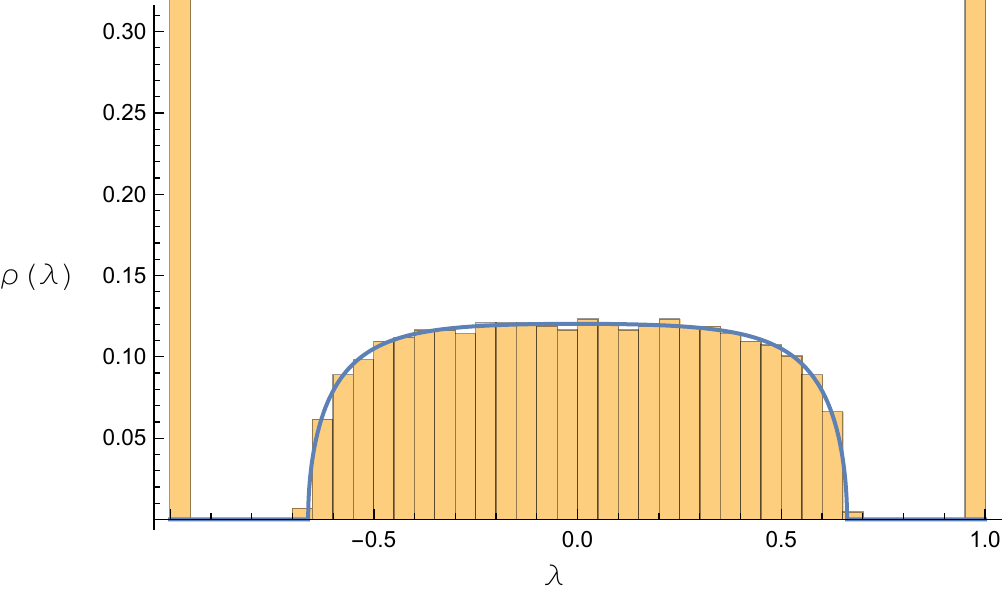} }\\
\multicolumn{2}{c}{ (e)}
\end{tabular}
\caption{Histograms of eigenvalues for different microcanonical truncation sizes of an originally 
$10000\times10000$ matrix ($2D=10000$). 
The blue curve represents the analytical prediction given by the Jacobi distribution of Eq.~(\ref{rholam}). 
Figs. (a)--(e) show results for $K/2D = 0.125$, 0.25, 0.5, 0.75, 0.875. 
Figs. (d) and (e) are truncated in height and do not show the full count of $\pm 1$ eigenvalues.}
\end{figure*}
\end{center}
\clearpage

\appendix
\section{Truncations Greater than Half}
\label{poles}
We write $A = U^\dagger \tilde A U$ in block-diagonal form,
\beq
A = \begin{pmatrix}
A_K & B \\
\noalign{\smallskip}
B^\dagger & A_{2D-K}
\end{pmatrix}.
\label{amat}
\eeq
We take $K\le 2D$. 
Since the eigenvalues of $A$ are $\pm 1$, we have
\beq
A^2 = I.
\label{A2I}
\eeq
This yields the three equations 
\begin{align} 
A_K^2+ BB^\dagger &= I,
\label{a2bb} \\
A_KB + BA_{2D-K} &= 0,
\label{abba} \\
B^\dagger\! B +A_{2D-K}^2 &= I.
\label{bba2}
\end{align} 
We can make independent unitary transformations on the upper $K\times K$ block
and on the lower $(2D{-}K)\times(2D{-}K)$ block that render $A_K$ and $A_{2D-K}$ diagonal.
We then write
\beq
(A_K)_{ij} =\lambda_i\delta_{ij}, \quad (A_{2D-K})_{i'j'} =\kappa_{i'}\delta_{i'j'},
\label{AKAD}
\eeq
where $i,j=1,\ldots,K$ and $i',j'=K{+}1,\ldots,2D{-}K$.
Taking the $ij'$ matrix element of Eq.~(\ref{abba}), we get
\beq
(\lambda_i + \kappa_{j'})B_{ij'}=0.
\label{lamkap}
\eeq
This shows that a nonvanishing matrix element of $B$ is possible if and only if 
there is an eigenvalue $\kappa_{j'}$ of $A_{2D-K}$ that is the negative of an
eigenvalue $\lambda_i$ of $A_K$. For $K<D$, there are more eigenvalues
of $A_{2D-K}$ than there are of $A_K$, and Eq.~(\ref{bba2}) then implies that
these extra eigenvalues must be $\pm1$. Since $\Tr A=0$, these extra eigenvalues
must come in $\pm 1$ pairs. 

We conclude that, for $K\le D$, the $2D{-}K$ eigenvalues of $A_{2D-K}$ consist of
$K$ eigenvalues that are equal in magnitude and opposite in sign to the $K$ eigenvalues
of $A_K$, $D{-}K$ eigenvalues $+1$, and $D{-}K$ eigenvalues $-1$.

\bibliographystyle{unsrt}
\bibliography{truncations.bib}
\end{document}